\documentclass[aps,prl,twocolumn,amsfonts,amssymb,amsmath,floats,floatfix,showpacs,preprintnumbers,superscriptaddress]{revtex4-1}

\usepackage{graphicx}
\usepackage{color}
\usepackage{dcolumn}
\usepackage{bm}
\usepackage{hyperref}
\usepackage{relsize}
\hypersetup{
colorlinks=true,
urlcolor= blue,
citecolor=blue,
linkcolor= blue,
bookmarks=true,
bookmarksopen=false,
}

\begin{document}

\title{Photoelectron spin-polarization-control in the topological insulator Bi$_2$Se$_3$}

\author{Z.-H. Zhu}
%\email{zhzhu@physics.ubc.ca}
\affiliation{Department of Physics {\rm {\&}} Astronomy, University of British Columbia, Vancouver, British Columbia V6T\,1Z1, Canada}
\author{C.N. Veenstra}
\affiliation{Department of Physics {\rm {\&}} Astronomy, University of British Columbia, Vancouver, British Columbia V6T\,1Z1, Canada}
\author{S. Zhdanovich}
\affiliation{Department of Physics {\rm {\&}} Astronomy, University of British Columbia, Vancouver, British Columbia V6T\,1Z1, Canada}
\author{M.P. Schneider}
\affiliation{Department of Physics {\rm {\&}} Astronomy, University of British Columbia, Vancouver, British Columbia V6T\,1Z1, Canada}
\author{T. Okuda}
\affiliation{Hiroshima Synchrotron Radiation Center, Hiroshima University, 2-313 Kagamiyama, Higashi-Hiroshima 739-0046, Japan}
\author{K. Miyamoto}
\affiliation{Hiroshima Synchrotron Radiation Center, Hiroshima University, 2-313 Kagamiyama, Higashi-Hiroshima 739-0046, Japan}
\author{S.-Y. Zhu}
\affiliation{Graduate\,School\,of\,Science,\,Hiroshima\,University,\,1-3-1\,Kagamiyama,\,Higashi-Hiroshima\,739-8526,\,Japan}
\author{\\H. Namatame}
\affiliation{Hiroshima Synchrotron Radiation Center, Hiroshima University, 2-313 Kagamiyama, Higashi-Hiroshima 739-0046, Japan}
\author{M. Taniguchi}
\affiliation{Hiroshima Synchrotron Radiation Center, Hiroshima University, 2-313 Kagamiyama, Higashi-Hiroshima 739-0046, Japan}
\affiliation{Graduate\,School\,of\,Science,\,Hiroshima\,University,\,1-3-1\,Kagamiyama,\,Higashi-Hiroshima\,739-8526,\,Japan}
\author{M.W. Haverkort}
\affiliation{Max Planck Institute for Chemical Physics of Solids, 01187 Dresden, Germany}
\affiliation{Quantum Matter Institute, University of British Columbia, Vancouver, British Columbia V6T\,1Z4, Canada}
\author{I.S. Elfimov}
\affiliation{Department of Physics {\rm {\&}} Astronomy, University of British Columbia, Vancouver, British Columbia V6T\,1Z1, Canada}
\affiliation{Quantum Matter Institute, University of British Columbia, Vancouver, British Columbia V6T\,1Z4, Canada}
\author{A. Damascelli}
\email{damascelli@physics.ubc.ca}
\affiliation{Department of Physics {\rm {\&}} Astronomy, University of British Columbia, Vancouver, British Columbia V6T\,1Z1, Canada}
\affiliation{Quantum Matter Institute, University of British Columbia, Vancouver, British Columbia V6T\,1Z4, Canada}

\date{\today}

\begin{abstract}
We study the manipulation of the photoelectron spin-polarization in Bi$_2$Se$_3$ by spin- and angle-resolved photoemission spectroscopy. General rules are established that enable controlling the spin-polarization of photoemitted electrons via light polarization, sample orientation, and photon energy. We demonstrate the $\pm$100\% reversal of a single component of the measured spin-polarization vector upon the rotation of light polarization, as well as a full three-dimensional manipulation by varying experimental configuration and photon energy. While a material-specific density-functional theory analysis is needed for the quantitative description, a minimal two-atomic-layer model qualitatively accounts for the spin response based on the interplay of optical selection rules, photoelectron interference, and topological surface-state complex structure. It follows that photoelectron spin-polarization control is generically achievable in systems with a layer-dependent, entangled spin-orbital texture.
\end{abstract}


\maketitle

The central goal in the field of spintronics is to realize highly spin-polarized electron currents and to actively manipulate their spin polarization direction. Topological insulators (TIs), as a new quantum phase of matter with a spin-polarized topologically-protected surface state \cite{Hasan:2010PRM, Qi:RMP, Moore:2010}, hold great promise for the development of a controllable `spin generator' for quantum spintronic applications \cite{Pesin:2012p8592}. A possible avenue is via the spin Hall effect and the spin currents that appear at the boundaries of TI systems, and the electric-field-induced magnetization switching achieved at the interface between a TI and a ferromagnet \cite{Franz:spintronic}. In addition, it has been demonstrated that a spin-polarized photocurrent can be generated from the topological surface state ({TSS}) using polarized light, suggesting the possibility of exploiting TIs as a material-platform for novel optospintronic devices \cite{Mciver:2011, PhysRevB.83.035309, Refael:CP}. 

All these exciting developments fundamentally rely on the spin properties of the TSS, which have been extensively studied by density functional theory (DFT) \cite{Zhang:2009BeSe,Yazyev:dft, Guo:dft} and spin- and angle-resolved photoemission spectroscopy (spin-ARPES) \cite{Hsieh:2009BiSe, Souma:spin, Xu:spin, Pan:spin, Jozwiak, Jozwiak:2013p8563, Cao:spin, xie:spin}. In Bi$_2$Se$_3$, we have shown that the TSS is not a simple two-dimensional state. Rather, it has a {\it layer-dependent spin-orbital entangled structure} -- extending over 10 atomic layers ($\sim\!2$\,nm) -- challenging the hypothesis of 100\% spin-polarization for the TSS Dirac fermions \cite{Zhu:BiSe_P}. Our DFT work also suggested a new pathway to control the spin polarization of photoelectrons via photon energy and linear polarization \cite{Zhu:BiSe_P}; although this is consistent with some experimental observations by spin-ARPES \cite{Jozwiak:2013p8563, Cao:spin, xie:spin}, no conclusive understanding of the phenomenon and its governing principles has yet been achieved. This is of critical importance for future applications, and will require a full examination of the photoelectron spin-polarization response in specifically designed spin-ARPES experiments.
\begin{figure}[b!]
\includegraphics[width=1\linewidth]{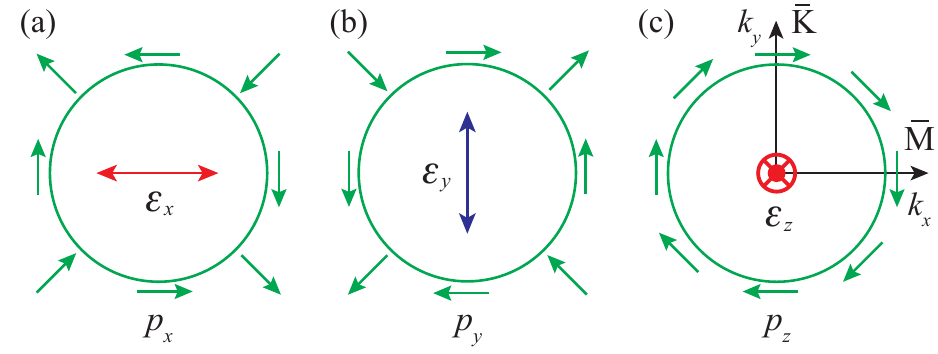}
\caption {\label{fig:spinSB_orbital} (color online). (a)--(c) In-plane spin texture as obtained separately for the $p_x$ (a), $p_y$ (b), and $p_z$ (c) orbital contributions to the topological surface state (TSS). Red/blue arrows indicate the light electric field ($\pi/\sigma$ polarization) that must be used to excite photoelectrons from each of the individual orbitals, according to the electric dipole selection rules.}
\end{figure}
\begin{figure}[t!]
\includegraphics[width=1\linewidth]{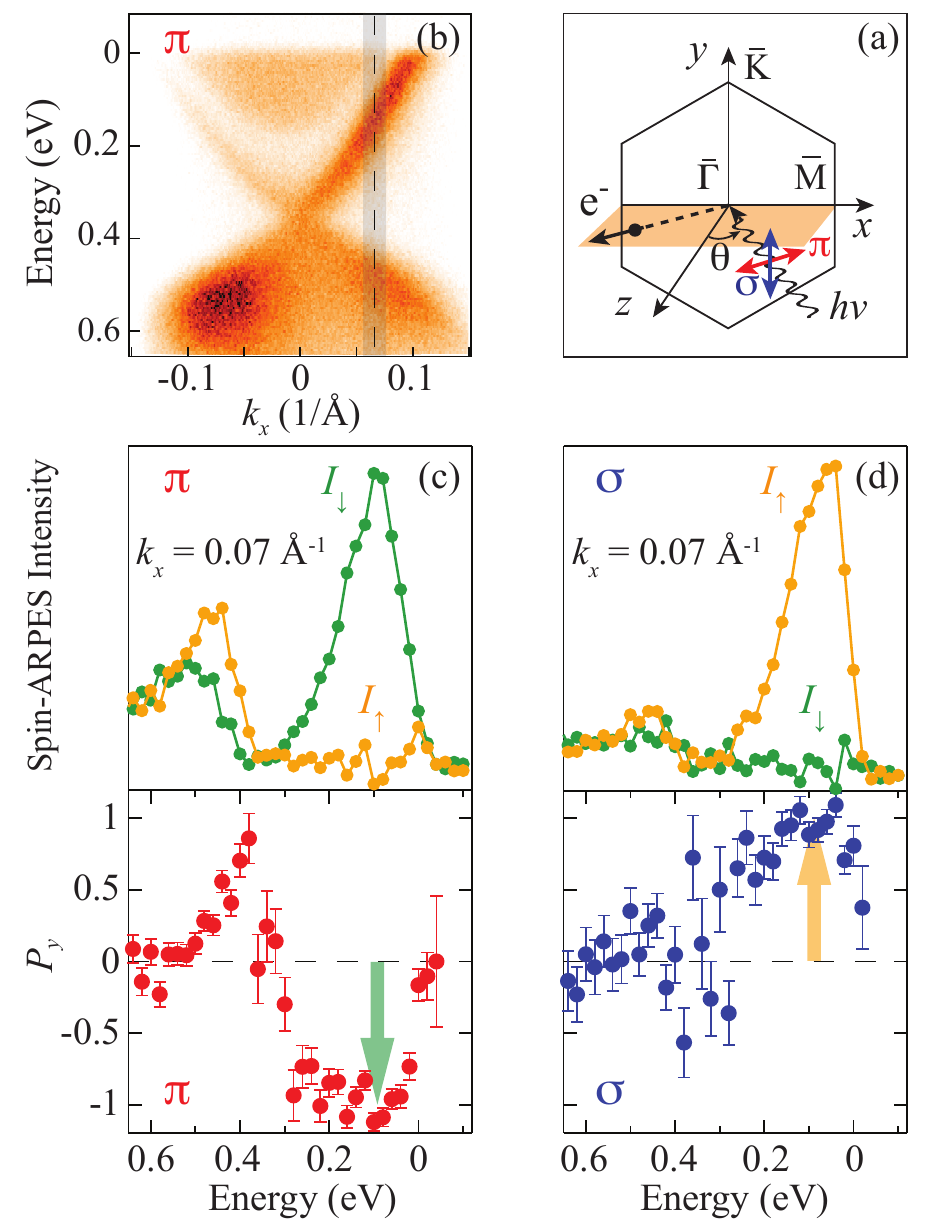}
\caption{\label{fig:spinSB_polar} (color online). (a) Schematics of the experimental geometry, with $\pi$ (horizontal) and $\sigma$ (vertical) linear polarization also indicated. (b) ARPES dispersion of TSS Dirac fermions measured along the $\bar M\!-\!\bar\Gamma\!-\!\bar M$ direction with $\pi$ polarization. (c) The top panel shows spin-ARPES EDCs, with spin quantization axis along the $y$ direction, measured with $\pi$ polarization along the gray-bar highlighted in (b) \cite{second}; the corresponding  $P_y$ spin polarization is shown in the lower panel (the TSS is located at 0.1\,eV in these data taken at $k_x\!=\!0.07\mathrm{\AA^{-1}}$). (d) Spin-ARPES data analogous to those in (c), now measured with $\sigma$ polarization.}
\end{figure}

In this Letter -- guided by a DFT analysis of the TSS entangled spin-orbital texture -- we present a systematic spin-ARPES study to elucidate the dependence of the photoelectron spin on light polarization, experimental geometry, and photon energy. We demonstrate a reversal of the spin polarization from $-$100\% to +100\% upon switching from $\pi$ to $\sigma$ polarized light. By changing sample geometry and tuning photon energy we can manipulate the photoelectron spin polarization in three dimensions. While a material-specific DFT analysis is needed for the complete quantitative description, here we introduce a minimal and fully-general two-atomic-layer model that qualitatively captures the unusual spin-ARPES response in terms of TSS spin-orbital texture, optical selection rules, and photoelectron interference. This paves the way to generating fully controllable spin-polarized photocurrents in TI-based optospintronic devices.

Spin-ARPES experiments were performed at the Hiroshima Synchrotron Radiation Center (HSRC) on the Efficient SPin REsolved Spectroscopy (ESPRESSO) endstation \cite{okuda:103302, Miyamoto:spin}, with 50\,meV and $\leq 0.04\,\mathrm{\AA^{-1}}$ energy and momentum resolution, respectively. This spectrometer can resolve both in-plane ($P_{x,y}$) and out-of-plane ($P_z$) photoelectron spin-polarization components. These are obtained from the relative difference between the number of spin-up and spin-down photoelectrons, according to the relation $P_{x, y, z}\!=\!(I^{\uparrow_{x, y, z}}\!-\!I^{\downarrow_{x, y, z}})/({I^{\uparrow_{x, y, z}}\!+\!I^{\downarrow_{x, y, z}}})$, and are presented in the sample frame. The crystals were oriented by Laue diffraction and cleaved in-situ at $\sim\!7\times 10^{-11}$\,torr; all measurements were performed at 30\,K once the surface evolution had mostly stabilized \cite{Zhu:RB}, using 21\,eV photons unless otherwise specified.

In Bi$_2$Se$_3$, the TSS wavefunction is composed of both in-plane ($p_{x, y}$) and out-of-plane ($p_z$) orbitals. As a consequence of spin-orbit coupling, the spin texture associated with each orbital is remarkably different, and has been referred to as {\it entangled spin-orbital texture} \cite{Zhu:BiSe_P, zhang:spinorbital,Cao:s}. In Fig.\,\ref{fig:spinSB_orbital}, we sketch the orbital-dependent in-plane spin polarization of the upper-branch Dirac fermions (with the out-of-plane spin component not shown). We see that the well-known TSS chiral spin texture is only present in the out-of-plane $p_z$ orbitals [Fig.\,\ref{fig:spinSB_orbital}(c)]; instead, the individual $p_{x}$ and $p_{y}$ spin configurations are not chiral, and are also opposite to one another [Figs.\,\ref{fig:spinSB_orbital}(a) and\,\ref{fig:spinSB_orbital}(b)]. By comparing the spin orientation of in-plane and out-of-plane orbitals, we learn that at different momentum-space locations they can be parallel, anti-parallel, or even perpendicular to each other. For example, $p_x$ and $p_z$ spin polarizations are parallel along the $\bar\Gamma\!-\!\bar M$ direction (i.e., the $k_x$ axis), but antiparallel along $\bar\Gamma\!-\!\bar K$ (i.e., the $k_y$ axis). As for probing these different orbital-dependent configurations, we note that  -- based on the optical selection rules and assuming excitations into final states of $s$ symmetry -- photoelectrons are emitted from a given $p_{x,y,z}$ orbital if the photon electric field has a non-zero component $\mathlarger{\mathlarger{\mathlarger{\varepsilon}}}_{x,y,z}$ along the corresponding direction \cite{Damascelli:physica}. Thus, using linearly polarized photons with electric field parallel to the $k_x$/$k_y$/$k_z$ directions, we can probe the $p_x$/$p_y$/$p_z$ spin textures individually in spin-ARPES (Fig.\,\ref{fig:spinSB_orbital}). 
\begin{figure*}[thp]
\includegraphics[width=1\linewidth]{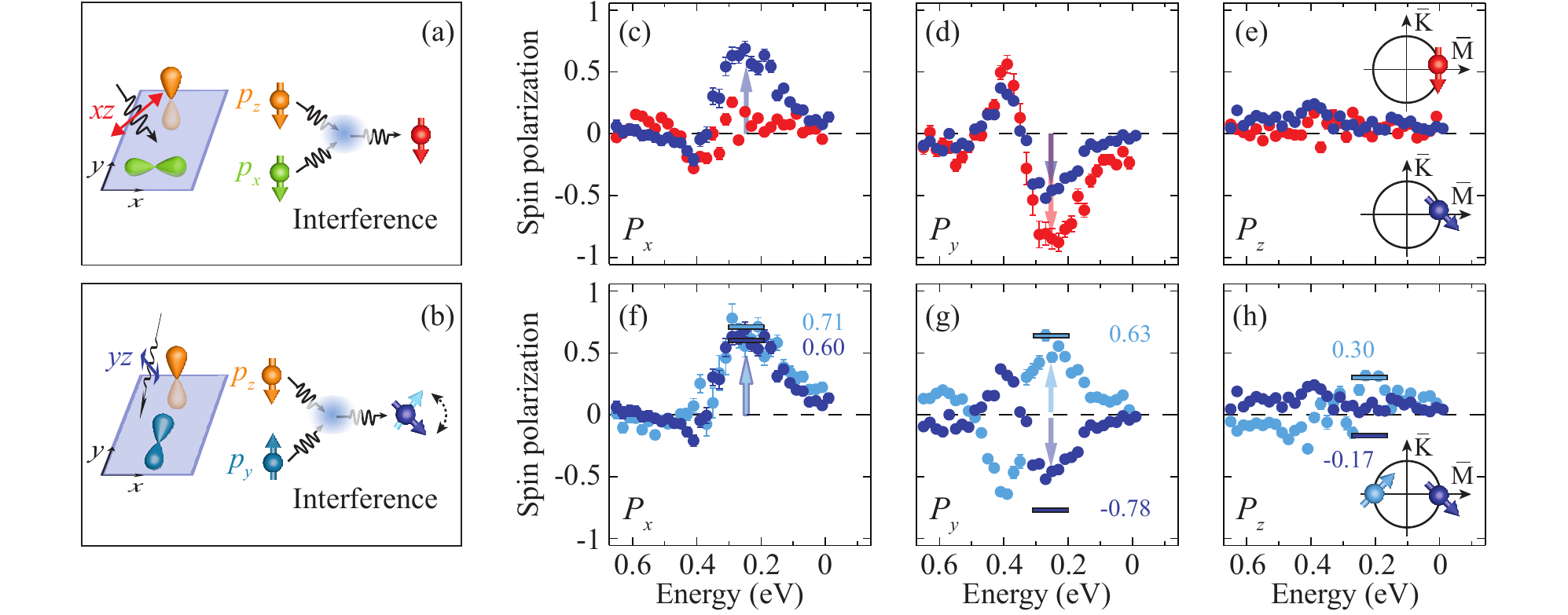}
\caption {\label{fig:spinSB_geo} (color online). (a),(b) Schematics of photoelectron interference effects in two configurations: (a) $\pi$-polarization incident in the $xz$ plane probes $p_x$ and $p_z$ orbitals with the same spin state (Fig.\,\ref{fig:spinSB_orbital}); (b) when incident in the $yz$ plane, $\pi$-polarization probes $p_y$ and $p_z$ orbitals with opposite spin states (Fig.\,\ref{fig:spinSB_orbital}). (c)--(e) Spin polarization curves at the $+k_x$ point as sketched in (e), measured for (a)/(b) configurations (red/blue curves). (f)--(h) Spin polarization curves at $\pm k_x$ as sketched in (h)  for the (b) configuration, together with $k_x\!=\!\pm0.04\,\mathrm{\AA}^{-1}$ DFT calculated values \cite{deviation} [for the red data in (c)--(e), DFT gives $\Vec{P}\!=\!(0,-1,0)$].}
\end{figure*}

Fig.\,\ref{fig:spinSB_polar} demonstrates the $\pm$100\% manipulation of photoelectron spin-polarization upon switching the light polarization from $\pi$ to $\sigma$ in spin-ARPES. When we measure the energy distribution curve (EDC) at $k_x\!=\!0.07\,\mathrm{\AA^{-1}}$ with $\pi$ polarization [photon electric field in the $xz$ plane, as in Figs.\,\ref{fig:spinSB_polar}(a,b)], we observe a peak only in the spin-down $y$-channel at the TSS upper-branch binding energy at $\sim\!0.1$\,eV [green curve in the top panel of Fig.\,\ref{fig:spinSB_polar}(c)]. Thus we obtain $P_{x,z}\!\simeq\!0$ [as shown by the red datasets in Figs.\,\ref{fig:spinSB_geo}(c,e)], and remarkably $P_y\!\simeq\!-$100\% for the spin-polarization vector components, as highlighted in the bottom panel of Fig.\,\ref{fig:spinSB_polar}(c) by the green arrow at 0.1\,eV (note that the positive $P_y$ value at $\sim\!0.4$\,eV originates from the TSS bottom branch and its reversed spin helicity \cite{Zhu:BiSe_P, zhang:spinorbital}). Most importantly, when light polarization is switched from $\pi$ to $\sigma$, while $P_{x,z}$ remain zero $P_y$ suddenly becomes +100\% at 0.1\,eV, as shown in Fig.\,\ref{fig:spinSB_polar}(d).

We note that a spin polarization as high as $\pm$100\% is rarely reported in previous spin-ARPES studies of Bi$_2$Se$_3$ \cite{Hsieh:2009BiSe, Souma:spin, Pan:spin, Xu:spin, Jozwiak, Jozwiak:2013p8563}; this is achieved in this study owing to the high efficiency of the spin polarimeter and the perfect alignment within the photoelectron emission plane of both the light polarization and sample $\bar\Gamma\!-\!\bar M$ direction, which eliminates the interference-induced deviations to be discussed below. The spin-polarization switching in Fig.\,\ref{fig:spinSB_polar} can be directly visualized based on the experimental configuration and the entangled spin-orbital texture (Fig.\,\ref{fig:spinSB_orbital}): $\pi$ polarization excites photoelectrons from $p_x$ and $p_z$ orbitals only, both of which are $-$100\% spin polarized along $y$ for all positive `$+k_x$' locations [Figs.\,\ref{fig:spinSB_orbital}(a,c)]; this gives $P_y\!\equiv\!-$100\% in spin-ARPES, consistent with the experiment in Fig.\,\ref{fig:spinSB_polar}(c). On the contrary, in $\sigma$ polarization photoelectrons originate only from the $p_y$ orbitals, which at $+k_x$ locations are +100\% spin-polarized along the $y$ direction, i.e. $P_y\!\equiv\!+$100\% as detected in Fig.\,\ref{fig:spinSB_polar}(d). 

By rotating light polarization between $\sigma$ and $\pi$, we would observe a continuous change of $P_y$ between $\pm$100\%, as experimentally verified by Jozwiak et al. \cite{Jozwiak:2013p8563}. Here we argue that, in addition to the TSS symmetry properties already accounted for in previous work \cite{Jozwiak:2013p8563,park:spin,wang}, also the TSS layer-dependent spin-orbital texture must be taken into account to fully explain the manipulation of photoelectron spin polarization by light, as evidenced by the dependence on geometry and photon-energy presented below. The spin-ARPES response is indeed most unusual for configurations different from the one in Fig.\,\ref{fig:spinSB_polar} -- which is unique in that electrons photoemitted by either $\pi$ or $\sigma$ light all have the same spin polarization even if originating from multiple orbitals.
\begin{figure*}[t!]
\includegraphics[width=1\linewidth]{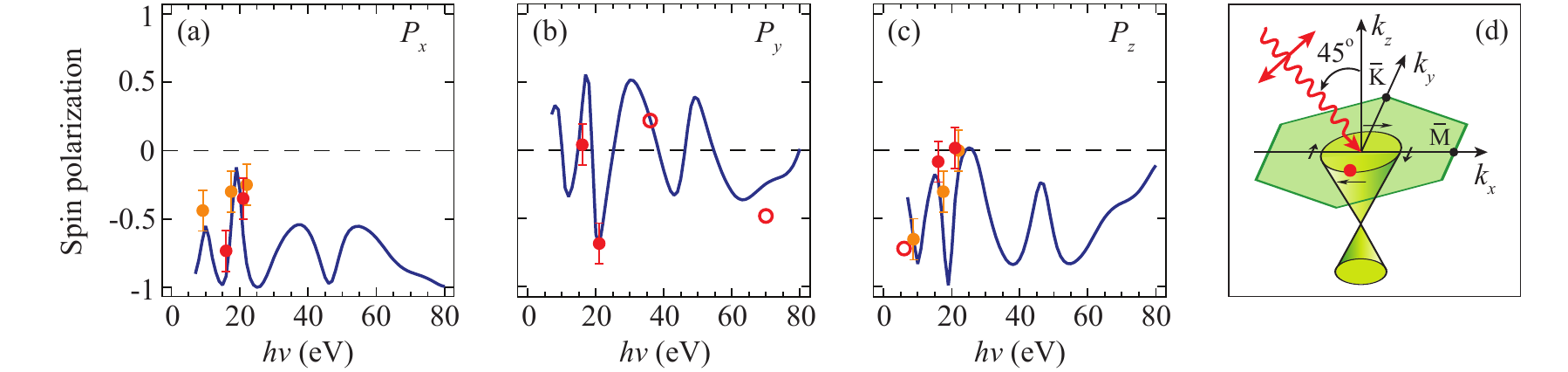}
\caption {\label{fig:spinSB_hv} (color online). (a)--(c) Solid blue lines: calculated photon-energy-dependence of the photoelectron spin-polarization-vector components, as obtained at the $-k_y$ point for $\pi$-polarized light incident in the $xz$ plane as shown in the sketch in (d). Solid red symbols are spin-ARPES data from this work; open red symbols are from Ref.\onlinecite{Jozwiak, Jozwiak:2013p8563}; solid yellow symbols were measured on crystals from Golden's group \cite{Golden:CuBiSe} and a 6-quintuple-layer film (22\,eV data) by Tjernberg and collaborators \cite{berntsen:instr}.}
\end{figure*}
This is shown in Figs.\,\ref{fig:spinSB_geo}(c--e) where we examine the photoelectron spin polarization at $+k_x$ \cite{fig3_kx}, for the two configurations of Figs.\,\ref{fig:spinSB_geo}(a,b). {\it Case I} -- $\mathlarger{\mathlarger{\mathlarger{\boldsymbol{\varepsilon}}}}\!\!\parallel\!\!xz$\,: photoelectrons are emitted from $p_{x,z}$ orbitals in the same spin state [Fig.\,\ref{fig:spinSB_geo}(a)], and as before we observe a close to $-100$\% $P_y$ \cite{sampleage} and zero $P_{x,z}$ [red symbols in Figs.\,\ref{fig:spinSB_geo}(c--e)]. {\it Case II} -- $\mathlarger{\mathlarger{\mathlarger{\boldsymbol{\varepsilon}}}}\!\!\parallel\!\!yz$\,: photoelectrons are emitted from $p_{y,z}$ orbitals with mixed spin states [Fig.\,\ref{fig:spinSB_geo}(b)], and are no longer fully polarized along $P_y$. Instead $P_y$ decreases and an unexpected -- within a two-dimensional TSS description --  $P_x\!\simeq\!70$\% appears [blue symbols in Figs.\,\ref{fig:spinSB_geo}(c--e) and sketch in \ref{fig:spinSB_geo}(e)]. Another interesting aspect is that while both $P_y$ and $P_z$ \cite{sign} switch sign at opposite momenta $\pm k_x$, as expected from time-reversal symmetry [Figs.\,\ref{fig:spinSB_geo}(g,h)], the $P_x$ retains the same non-zero value  [Figs.\,\ref{fig:spinSB_geo}(f) and sketch in \ref{fig:spinSB_geo}(h)].

To understand the unexpected results of Fig.\,\ref{fig:spinSB_geo} -- seemingly inconsistent with the TSS time-reversal invariance  -- we need to consider photoelectron-interference effects specific for spin-ARPES. To this end, we express the measured spin polarization vector $\Vec{P}$ in terms of the expectation value of generalized spin operators \cite{supplementary}:
\begin{equation}
\begin{split}
P_{\eta} = & \frac{\sum_{i, \tau}   \langle S_{\eta}^{i, \tau;\,i, \tau } \rangle |M_{i,\tau}|^{2}}{I_{total}}  \\
&\hspace{-0.29cm}+ \frac{\sum_{i \neq i' \!,\, \tau \neq \tau'}  \langle S_{\eta}^{i, \tau;\, i', \tau'} \rangle \,e^{\mathrm{i} k_{z}(z_{i}-z_{i'})} M_{i, \tau}^{\ast}M_{i', \tau'}}{I_{total}},
\label{eq:SpinBS}
\end{split}
\end{equation}
where $\eta\!\in\!\{x, y, z\}$, $\tau\!\in\!\{p_x,p_y,p_z\}$, $i$ is the atomic-layer index (the TSS layer-dependent structure is a key factor here \cite{Zhu:BiSe_P}); $M_{i, \tau}\propto\langle e^{\mathrm{i} \mathrm{\bf k_{\parallel}}  \cdot \mathrm{\bf r_{\parallel}}} | \mathrm{\bf A} \cdot \mathrm{\bf p}  | \psi_{i,\tau} \rangle$ is the matrix element of the optical transition between an atomic wavefunction of orbital $\tau$ centered around the atomic layer $i$ and a free-electron final state; the $k_z$ part of the latter has been factorized in the phase term $e^{\mathrm{i} k_{z}(z_{i}-z_{i'})}$, which accounts for the optical path difference for photoelectrons from different layers; and $I_{total}$ is the sum of intensity from spin-up and spin-down channels. The generalized spin operator in the expectation value $\langle S_{\eta}^{i, \tau;\, i', \tau'}\!\rangle$ is defined as:
\begin{equation}
S_{\eta}^{i, \tau;\, i', \tau'} = |\psi_{i,\tau} \rangle \langle \psi_{i', \tau'} | \sigma_{\eta},
\label{eq:Soperator}
\end{equation}
where $\sigma_{x,y,z}$ are the Pauli spin matrices. The crucial point is that in  Eq.\,\ref{eq:SpinBS} the $i \neq i' \!,\, \tau \neq \tau'$ {\it off-diagonal terms account for the interference effects}. If the initial states $\psi_{i,\tau}$ and $\psi_{i', \tau'}$ being probed all have the same spin expectation value, then $\langle S_{\eta}^{i, \tau;\, i, \tau}\!\rangle\!=\!\langle S_{\eta}^{i, \tau;\, i', \tau'}\!\rangle$ and $P_{\eta}\!=\!100$\% for the $\eta$ component corresponding to the spin quantization axis, as in {\it Case I} of Fig.\,\ref{fig:spinSB_geo}(a). However, when the initial states being simultaneously probed have different spin states, as in {\it Case II} of Fig.\,\ref{fig:spinSB_geo}(b), non-trivial effects should be expected for the measured spin polarization due to the contribution of the $S_{\eta}^{i, \tau;\, i', \tau '}$ {\it interference term}.

To qualitatively demonstrate that Eq.\,\ref{eq:SpinBS} describes the spin-ARPES results in Fig.\,\ref{fig:spinSB_geo}, in the Supplemental Material we build a phenomenological two-atomic-layer wavefunction as the minimal model needed to capture interference effects \cite{supplementary}, starting from the effective TSS wavefunction derived by Zhang {\it et al.} \cite{Zhang:2009BeSe,Liu:model, zhang:spinorbital}. For the upper branch of the Dirac-cone this becomes \cite{supplementary}:
\begin{equation}
\Psi\!=\!\!\sum_{i=1}^{2} \frac{\alpha_{i}}{\sqrt{2}} {\mathrm{i} e^{-\mathrm{i} \varphi} \choose 1} |p_{z}\rangle\!-\!\frac{\beta_{i}}{2}{  -1 \choose \mathrm{i} e^{-\mathrm{i} \varphi}} |p_{x}\rangle\!+\!\frac{\beta_{i}}{2} {- \mathrm{i} \choose e^{-\mathrm{i} \varphi}} |p_{y}\rangle
\label{eq:psi3md}
\end{equation}
where $\alpha_{i}$ and $\beta_{i}$ are layer-dependent coefficients, and the in-plane phase $\varphi$ (defined as the angle between \textbf{k} and the $+k_x$ direction) reproduces the orbital-dependent spin texture shown in Fig.\,\ref{fig:spinSB_orbital}. To further simplify the problem we assume -- without loss of generality -- that $\alpha_1\!=\!\beta_2\!=\!0$, $\alpha_2\!=\!\sqrt{3/2}$, and $\beta_1\!=\!1/\sqrt{2}$; this choice matches the 1:3 overall in-plane/out-of-plane orbital weight ratio calculated by DFT for  Bi$_3$Se$_2$  \cite{Zhu:BiSe_P}. Then, for $\mathlarger{\mathlarger{\mathlarger{\boldsymbol{\varepsilon}}}}\!\!\parallel\!\!yz$ ({\it Case II}\,), the initial-state components being probed  reduce to \cite{supplementary}:
\begin{equation}
\Psi_{p_z} = {\frac{\sqrt3}{2}}{\mathrm{i} e^{-\mathrm{i} \varphi} \choose 1}\,\,\, \text{and}\,\,\,  \Psi_{p_y} =  \frac{\sqrt{2}}{4} {- \mathrm{i} \choose e^{-\mathrm{i} \varphi}}.
\label{eq:wavepzpy}
\end{equation}
At $\pm\!k_x$ ($\varphi\!=\!0$ and $\pi$, respectively), the intrinsic spin polarization is $\mp$100\% ($\pm\!100$\%) along the $k_y$ direction for the $p_z$ ($p_y$) orbital \cite{supplementary}, as in Fig.\,\ref{fig:spinSB_orbital}. By means of Eq.\,\ref{eq:SpinBS}, we can now calculate the photoelectron spin-polarization vector $\Vec{P}$ as seen at $\pm k_x$ in spin-ARPES, obtaining \cite{supplementary}:
\begin{equation}
\Vec{P} ({\pm k_x}) \propto (\sin\theta_{k_z}, \mp 0.6, \mp \cos\theta_{k_z}),
\label{eq:pmodel}
\end{equation}
where $\theta_{k_z} = k_z(z_1\!-\!z_2)$. We see that, although the spin polarization of each individual initial state is purely along $y$, the photoelectron spin polarization can have non-zero components along $x$ and/or $z$, if $z_1-z_2\!\neq\!0$. This highlights the need for a minimal two-atomic-layer model. Also note that the explicit presence of $k_z$ leads to photon-energy-dependence (more below), and all $P_{x,y,z}$  components oscillate sinusoidally with different phases, upon varying $k_z$ \cite{supplementary}; this is responsible for the  maximal $P_{x}$ and minimal  $P_{z}$ in Figs.\,\ref{fig:spinSB_geo}(f)--\ref{fig:spinSB_geo}(h). Finally, Eq.\,\ref{eq:pmodel} confirms that only $P_y$ and $P_z$ components reverse their signs, while $P_x$ retains the same value at $\pm k_x$, again as observed in our spin-ARPES data in Figs.\,\ref{fig:spinSB_geo}(f)--\ref{fig:spinSB_geo}(h) \cite{complex}.

While our two-atomic-layer model reproduces the spin-ARPES results qualitatively, we stress that the quantitative description must be based on the complete $\sim$10-atomic-layer TSS wavefunction obtained for Bi$_2$Se$_3$ by DFT \cite{Zhu:BiSe_P}. To this end, in Fig.\,\ref{fig:spinSB_hv} we present the photon-energy-dependence of the photoelectron spin polarization  $P_{x,y,z}$  at $-k_y$, for $\mathlarger{\mathlarger{\mathlarger{\boldsymbol{\varepsilon}}}}\!\!\parallel\!\!xz$. We find that our DFT-based results -- with their remarkable oscillating behavior, which however always guarantees $|\Vec{P}|\!=\!1$ -- are in agreement with the spin-ARPES data from this and other studies \cite{Jozwiak, Jozwiak:2013p8563}. This  conclusively demonstrates that the photon-energy-controlled photoelectron spin polarization stems from interference effects acting in concert with the TSS layer-dependent, entangled spin-orbital texture.

In conclusion, we have explained the underlying mechanism of the manipulation of photoelectron spin polarization in TIs, as a consequence of the TSS entangled spin-orbital texture, optical selection rules, and quantum interference. This is responsible also for the significantly different ARPES intensities observed at $\pm\!k_{x}$ in Fig.\,\ref{fig:spinSB_polar}(b), implying that a net spin-polarized current can be photoinduced by linearly polarized light \cite{Mciver:2011}. Thus, our spin-ARPES study demonstrates how to generate a spin-polarized photocurrent in Bi$_2$Se$_3$ and manipulate its absolute spin polarization by linearly polarized light, a key step in TI-based optospintronics. We argue that all these phenomena could be valid in other spin-orbit coupled systems, as long as the initial states are characterized by a layer-dependent entangled spin-orbital texture.

We thank M. Franz for discussions; T. Natsumeda and T. Warashina for technical assistance; M.H. Berntsen, O. G\"otberg, Y.K. Huang, B. Wojek, M.S. Golden, and O. Tjernberg for making available to us -- prior to publication -- the spin-ARPES data from both Bi$_2$Se$_3$ thin-films and single crystals indicated in yellow in Fig.\,\ref{fig:spinSB_hv}. This work was supported by the Max Planck - UBC Centre for Quantum Materials, the Killam, Alfred P. Sloan, Alexander von Humboldt, and NSERC's Steacie Memorial Fellowships (A.D.), the Canada Research Chairs Program (A.D.), NSERC, CFI, and CIFAR Quantum Materials. The experiments were performed with the approval of the Proposal Assessing Committee of the Hiroshima Synchrotron Radiation Center (Proposal No. 13-A-34).

%\vspace{-0.6cm}
\bibliography{BiSe_spin_second_revision}

\begin{thebibliography}{38}
\bibitem [{sec()}]{second}%
  \BibitemOpen
  \href@noop {} {\bibinfo  {journal} {These data were obtained in the second
  Brillouin zone (BZ) with a $-10^{\circ}$ light incidence angle; measurements
  in the first BZ with an angle of 50$^{\circ}$ gave similar results}\
  }\BibitemShut {NoStop}%
\bibitem [{fig()}]{fig3_kx}%
  \BibitemOpen
  \href@noop {} {\bibinfo  {journal} {In this experiment, performed along the
  $\bar\Gamma\!-\!\bar M$ direction at $|k_x|\!=\!0.04\,\mathrm{\AA^{-1}}$, the
  TSS is located at 0.25\,eV below $E_F$ (or, equivalently, at 0.1\,eV above
  the Dirac point)}\ }\BibitemShut {NoStop}%
\bibitem [{sam()}]{sampleage}%
  \BibitemOpen
\bibfield  {journal} {  }\href@noop {} {\bibinfo  {journal} {The slight
  reduction from $P_y\!=\!-100$\% measured initially is due to the sample
  surface degradation during the three continuous days of experiments after
  cleaving}\ }\BibitemShut {NoStop}%
\bibitem [{sig()}]{sign}%
  \BibitemOpen
\bibfield  {journal} {  }\href@noop {} {\bibinfo  {journal} {Although
  $P_z\!=\!0$ for the TSS upper branch, the sign change of $P_z$ can be
  observed for the TSS lower branch}\ }\BibitemShut {NoStop}%
\bibitem [{com()}]{complex}%
  \BibitemOpen
  \href@noop {} {\bibinfo  {journal} {The DFT $P_{x,y,z}$ values are asymmetric
  at $\pm k_x$ because $\alpha$ and $\beta$ in Eq.\,3 become complex numbers
  away from $\Gamma$}\ }\BibitemShut {NoStop}%
\end{thebibliography}

\end{document}